\documentclass[pra,twocolumn,notitlepage,superscriptaddress,nofootinbib,longbibliography]{revtex4-1}
\usepackage[english]{babel}
\usepackage{amsmath,amssymb,bbm,mathrsfs,color,graphicx,bm,braket,comment}
\usepackage[colorlinks,citecolor=blue,urlcolor=blue]{hyperref}

\newcommand{\Z}{\mathbbm{Z}}
\newcommand{\C}{\mathbbm{C}}

\newcommand{\bj}{\mathbf{j}}

\newcommand{\abs}[1]{\left\lvert #1 \right\rvert}

\newcommand{\unitvec}[1]{\hat{\mathbf{#1}}}

\begin{document}




\title{Lattice duality for the compact Kardar-Parisi-Zhang equation}

\author{L. M. Sieberer}

\affiliation{Department of Condensed Matter Physics, Weizmann Institute of
  Science, Rehovot 7610001, Israel}

\affiliation{Department of Physics, University of California, Berkeley,
  California 94720, USA}

\author{G. Wachtel}

\affiliation{Department of Condensed Matter Physics, Weizmann Institute of
  Science, Rehovot 7610001, Israel}

\affiliation{Department of Physics, University of Toronto, Toronto, Ontario M5S
  1A7, Canada}

\author{E. Altman}

\affiliation{Department of Condensed Matter Physics, Weizmann Institute of
  Science, Rehovot 7610001, Israel}

\affiliation{Department of Physics, University of California, Berkeley,
  California 94720, USA}

\author{S. Diehl}

\affiliation{Institute of Theoretical Physics, University of Cologne, D-50937
  Cologne, Germany}

\begin{abstract}
  A comprehensive theory of the Kosterlitz-Thouless transition in
  two-dimensional superfluids in thermal equilibrium can be developed within a
  dual representation which maps vortices in the superfluid to charges in a
  Coulomb gas. In this framework, the dissociation of vortex-antivortex pairs at
  the critical temperature corresponds to the formation of a plasma of free
  charges. The physics of vortex unbinding in driven-dissipative systems such as
  fluids of light, on the other hand, is much less understood. Here we make a
  crucial step to fill this gap by deriving a transformation that maps the
  compact Kardar-Parisi-Zhang (KPZ) equation, which describes the dynamics of
  the phase of a driven-dissipative condensate, to a dual electrodynamic
  theory. The latter is formulated in terms of modified Maxwell equations for
  the electromagnetic fields and a diffusion equation for the charges
  representing vortices in the KPZ equation. This mapping utilizes an adaption
  of the Villain approximation to a generalized Martin-Siggia-Rose functional
  integral representation of the compact KPZ equation on a lattice.
\end{abstract}

\maketitle



\section{Introduction}
\label{sec:introduction}

The Kardar-Parisi-Zhang (KPZ) equation~\cite{Kardar1986} represents a paradigm
of non-equilibrium statistical mechanics, describing universal scaling behavior
in a rich variety of physical systems. To name just a few examples, the range of
its applications includes the growth of bacterial
colonies~\cite{Vicsek1990,Wakita1997,Huergo2010}, fluid flow in porous
media~\cite{Rubio1989}, combustion of
paper~\cite{Maunuksela1997,Myllys2001,Miettinen2005}, and turbulent liquid
crystals~\cite{Takeuchi2010,Takeuchi2011}. Recently, it has been noted that the
KPZ equation emerges also in the context of condensation phenomena out of
thermodynamic equilibrium, in systems such as
exciton-polaritons~\cite{Carusotto2013,Byrnes2014}. The latter are bosonic
quasiparticles, formed via hybridization of photons in a semiconductor
microcavity and excitons confined in a two-dimensional quantum well, and have a
finite lifetime due to the leakage of the light field out of the
cavity. Compensating these losses by continuously injecting energy in the form
of laser light into the system drives it into a non-equilibrium steady state
that exhibits signatures of Bose-Einstein condensation of
exciton-polaritons~\cite{Carusotto2013,Byrnes2014}. Fluctuations of the phase of
such a condensate obey the KPZ
equation~\cite{Altman2015,He2015,Gladilin2014,Ji2015,Sieberer2015a}. The
fundamental difference to the above-mentioned cases of KPZ dynamics is that the
phase of the condensate is a compact variable and may thus contain topological
defects, i.e., vortices. In fact, driven-dissipative condensates are by far not
the only instance of compact KPZ dynamics: further examples include driven
vortex lattices in disordered superconductors~\cite{Aranson1998}, active
smectics~\cite{Chen2013}, and the phase dynamics of other systems obeying the
complex Ginzburg-Landau equation (CGLE) with noise~\cite{Aranson2002}; moreover,
a KPZ-type non-linearity occurs also in sliding charge-density
waves~\cite{Balents1995,Chen1996} and arrays of coupled limit-cycle
oscillators~\cite{Lauter2015}. This raises the question, how topological defects
can be incorporated systematically in the compact KPZ equation, and calls for a
formulation of the problem that treats these defects explicitly as fundamental
degrees of freedom of the system.

In thermal equilibrium, such a description of the physics in terms of
topological defects can be obtained by performing a duality transformation which
maps the partition function from a functional integral over a compact field to
one that is taken over configurations of defects. For systems defined on a
lattice, the dual description can be derived systematically using the Villain
approximation~\cite{Villain1975}. A case in point is the duality transformation
for the classical $XY$-model, which can be mapped to a Coulomb gas with charges
representing
vortices~\cite{Jose1977,Jose1978,Chaikin1995,Savit1980,Fisher1999}. The
topological nature of vortices is reflected in the quantization of the charges
to integer values. This vortex-charge duality has been extended to a
comprehensive theory describing also the dissipative dynamics of superfluid
films at finite temperature~\cite{Ambegaokar1978,Ambegaokar1980,Minnhagen1987},
and to a full quantum electrodynamics theory at zero
temperature~\cite{Fisher1989a}. Moreover, the duality transformation for Abelian
(and non-Abelian) lattice systems has been put on a systematic footing by making
use of Bianchi identities~\cite{Batrouni1984}.

In this paper, we derive a systematic lattice duality transformation for compact
KPZ dynamics. This derivation complements the heuristic derivation of a dual
electrodynamic theory in the continuum, which we introduced in
Ref.~\cite{Wachtel2016}. There, the dual theory served as the basis for a
detailed RG analysis of vortex unbinding in the non-equilibrium steady state.

The rest of the presentation is organized as follows: in
Sec.~\ref{sec:dual-transf-latt}, we present the derivation of the duality
transformation. In particular, we derive the Martin-Siggia-Rose (MSR) action for
the compact KPZ equation in Sec.~\ref{sec:msr-action-compact}, and describe the
appropriate form of the Villain approximation in
Sec.~\ref{sec:dual-transf}. There we also discuss how charges (vortices) can be
introduced as independent degrees of freedom by means of the Poisson summation
formula. In Sec.~\ref{sec:dual-electr-theory}, we show that the MSR functional
integral resulting from the duality transformation can be reduced to the dual
electrodynamics theory of Ref.~\cite{Wachtel2016}, which is formulated in terms
of Langevin equations for the electromagnetic fields and the charges. We finish
by giving in Sec.~\ref{sec:outlook} an outlook on possible applications of the
formalism developed in this paper. Technical details of the Villain
transformation are deferred to the Appendix.

\section{Duality transformation on a lattice}
\label{sec:dual-transf-latt}

In this section we present our main result, which is the derivation of the
lattice duality for the compact KPZ equation. As indicated in the introduction,
in thermal equilibrium, the duality transformation can be performed conveniently
for systems which are defined on a lattice. The key step is then to replace the
periodic potential of the compact field in the (functional integral
representation of the) partition function by a simplified Villain
form~\cite{Villain1975,Chaikin1995,Savit1980}.

Thus, to obtain the appropriate generalization of the Villain approximation to
the case of the compact KPZ equation, we begin by formulating the latter for a
lattice system. Then, in Sec.~\ref{sec:msr-action-compact}, we develop a
modification of the Martin-Siggia-Rose (MSR) functional integral for the lattice
compact KPZ equation. This modification of the usual MSR functional
integral~\cite{Kamenev2011,Altland2010a,Tauber2014a} is necessary in order to
properly account for the compactness of the phase field. For a classical system
out of thermal equilibrium, the MSR functional integral is the natural
counterpart to the partition function. Hence, it provides the appropriate
framework for performing a generalized Villain approximation in
Sec.~\ref{sec:dual-transf}.

\subsection{MSR action for the compact KPZ equation}
\label{sec:msr-action-compact}

Our starting point is the compact KPZ equation for the phase $\theta$ of a
two-dimensional driven-dissipative
condensate~\cite{Altman2015,He2015,Gladilin2014,Ji2015,Sieberer2015a}. In
spatial continuum, the compact KPZ equation takes the well-known form
\begin{equation}
  \label{eq:KPZ}
  \partial_t \theta = D \nabla^2 \theta + \frac{\lambda}{2} \left( \nabla \theta
  \right)^2 + \eta,
\end{equation}
where --- in contrast to the usual KPZ equation~\cite{Kardar1986} --- $\theta$
is defined on a circle, i.e., it takes values in the interval
$\theta \in [0, 2 \pi)$. $\eta$ is a Gaussian noise source with zero mean and
white spectrum,
\begin{equation}
  \label{eq:eta}
  \langle \eta(t, \mathbf{x}) \eta(t', \mathbf{x}') \rangle = 2 \Delta \delta(t
  - t') \delta(\mathbf{x} - \mathbf{x}').
\end{equation}
A systematic way to derive the KPZ equation on a lattice is to start from the
CGLE on a lattice (where spatial derivatives are replaced by standard hopping
terms) and integrate out density fluctuations as in the continuum
case~\cite{Altman2015,He2015,Gladilin2014,Ji2015,Sieberer2015a}. This amounts to
replacing spatial derivatives in the KPZ equation~\eqref{eq:KPZ} with finite
differences according to
\begin{equation}
  \label{eq:6}
  \begin{split}
    \nabla^2 \theta & \to - \sum_{\unitvec{a}} \sin(\theta_{\mathbf{x}} -
    \theta_{\mathbf{x} + \unitvec{a}}), \\ \left( \nabla \theta \right)^2 & \to
    - \sum_{\unitvec{a}} \left( \cos(\theta_{\mathbf{x}} - \theta_{\mathbf{x} +
        \unitvec{a}}) - 1 \right),
  \end{split}
\end{equation}
where $\mathbf{x} + \unitvec{a}$ are the nearest neighbors of the lattice site
$\mathbf{x}$, i.e., the sums are over
$\unitvec{a} \in \{ \pm \unitvec{x}, \pm \unitvec{y} \}$ (for convenience we
choose the lattice spacing as $a = 1$). Accordingly, the compact KPZ equation
reads
\begin{multline}
  \label{eq:Villain-3}
  \partial_t \theta_{\mathbf{x}} = - \sum_{\unitvec{a}} \left[ D
    \sin(\theta_{\mathbf{x}} - \theta_{\mathbf{x} + \unitvec{a}})
    \vphantom{\frac{\lambda}{2}} \right. \\ 
  \left. + \frac{\lambda}{2} \left( \cos(\theta_{\mathbf{x}} -
      \theta_{\mathbf{x} + \unitvec{a}}) - 1 \right) \right] +
  \eta_{\mathbf{x}},
\end{multline}
For $\lambda = 0$, this equation reduces to the form of the two-dimensional
classical $XY$-model with relaxational dynamics. Then, the Langevin equation can
be written as
$\partial_t \theta_{\mathbf{x}} = - \Gamma \delta
\mathcal{H}_{\mathit{XY}}/\delta \theta_{\mathbf{x}} + \eta_{\mathbf{x}}$,
where the $XY$-Hamiltonian reads (the sum is over pairs of neighboring sites)
\begin{equation}
  \label{eq:1}
  \mathcal{H}_{XY} = K \sum_{\langle \mathbf{x}, \mathbf{x}' \rangle}
  \cos(\theta_{\mathbf{x}} - \theta_{\mathbf{x}'}).
\end{equation}
Consequently, $D = \Gamma K$ in Eq.~\eqref{eq:Villain-3} is the product of the
diffusion constant $\Gamma$ and the spin-stiffness $K$. The stationary state of
the Fokker-Planck equation corresponding to relaxational Langevin dynamics is
the thermal Gibbs ensemble with distribution function
$\mathcal{P}_{\mathrm{Gibbs}} \propto \exp(- \mathcal{H}_{XY}/T)$ at temperature
$T = \Delta/\Gamma$. This is the starting point for deriving the dual Coulomb
gas representation of the equilibrium
$XY$-model~\cite{Jose1977,Jose1978,Chaikin1995,Savit1980,Fisher1999}. However,
in the KPZ problem, the closed form of the \emph{stationary} distribution is not
known in more than one dimension, and therefore we derive the dual
representation in terms of the \emph{dynamical} MSR functional instead.

Hence, the next step is to rewrite the Langevin equation~\eqref{eq:Villain-3} in
the form of an equivalent MSR action. We slightly modify the usual
approach~\cite{Kamenev2011,Altland2010a,Tauber2014a} in order to account for the
compactness of the phase: as customary, we discretize the stochastic process
described by Eq.~\eqref{eq:Villain-3} in time, i.e., we replace the continuous
function of time $\theta_{\mathbf{x}}(t)$ by a sequence $\theta_{t, \mathbf{x}}$
corresponding to specific points in time $t$ which are multiples of the temporal
lattice spacing $\epsilon$. Since each $\theta_{t, \mathbf{x}}$ is the phase of
a complex number $\psi_{t, \mathbf{x}}$, the discrete stochastic process has to
be invariant under shifts
$\theta_{t, \mathbf{x}} \mapsto \theta_{t, \mathbf{x}} + 2 \pi n_{t,
  \mathbf{x}}$
for integer-valued $n_{t, \mathbf{x}}$ (this is a gauge symmetry in the most
general sense of a redundancy of the description that is inherent to the polar
representation of the complex number $\psi_{t, \mathbf{x}}$). This property is
ensured if we write the update of $\theta_{t, \mathbf{x}}$ from time $t$ to
$t + \epsilon$ in the following way:
\begin{equation}
  \label{eq:Villain-78}
  \theta_{t + \epsilon, \mathbf{x}} = \theta_{t, \mathbf{x}} + \epsilon \left(
    \mathcal{L}[\theta]_{t, \mathbf{x}} + \eta_{t, \mathbf{x}} \right) + 2 \pi
  n_{t, \mathbf{x}}, 
\end{equation}
where by $\mathcal{L}[\theta]$ we denote the deterministic part of the compact
KPZ equation~\eqref{eq:Villain-3}, i.e.,
\begin{multline}
  \label{eq:Villain-6}
  \mathcal{L}[\theta]_{t, \mathbf{x}} = - \sum_{\unitvec{a}} \left[ D
    \sin(\theta_{t, \mathbf{x}} - \theta_{t, \mathbf{x} + \unitvec{a}})
    \vphantom{\frac{\lambda}{2}} \right. \\ \left. + \frac{\lambda}{2} \left(
      \cos(\theta_{t, \mathbf{x}} - \theta_{t, \mathbf{x} + \unitvec{a}}) - 1
    \right) \right].
\end{multline}
In Eq.~\eqref{eq:Villain-78}, $2 \pi n_{t, \mathbf{x}}$ is the unique multiple
of $2 \pi$ that has to be added to
$\theta_{t, \mathbf{x}} + \epsilon \left( \mathcal{L}[\theta]_{t, \mathbf{x}} +
  \eta_{t, \mathbf{x}} \right)$
so that the sum is in the interval from $0$ to $2 \pi$. Thus,
Eq.~\eqref{eq:Villain-78} defines a stochastic process for which
$\theta_{t, \mathbf{x}}$ remains within this interval at all times. Note that
because $n_{t, \mathbf{x}}$ is integer-valued, the straightforward continuum
limit $\epsilon \to 0$ of this stochastic process is ill-defined. However, below
we perform a sequence of manipulations leading eventually to a form which allows
us to take this limit.

In the following we use the symbol $\Delta_t$ to denote discrete derivatives
with respect to time, i.e., we write
$\Delta_t \theta_{t, \mathbf{x}} = \theta_{t + \epsilon, \mathbf{x}} -
\theta_{t, \mathbf{x}} \approx \theta_{t, \mathbf{x}} - \theta_{t - \epsilon,
  \mathbf{x}}$;
the form after the second equality appears below when we perform summations by
parts and we do not make a distinction between the two forms of the discrete
derivative as they are equivalent in the limit $\epsilon \to 0$.

We proceed with the construction of the MSR functional integral for
Eq.~\eqref{eq:Villain-78} in the usual
way~\cite{Kamenev2011,Altland2010a,Tauber2014a}. The solution of the stochastic
process Eq.~\eqref{eq:Villain-78} for a given realization of the noise is
denoted by $\theta_{\eta}$, and an arbitrary observable, which is a functional
of $\theta$, by $O[\theta]$.  Calculating the expectation value of $O[\theta]$
requires us to take the average of $O[\theta_{\eta}]$ over different noise
realizations, weighted by the Gaussian distribution function
\begin{equation}
  \label{eq:Villain-7}
  \mathcal{P}[\eta] \propto e^{- \frac{\epsilon}{4 \Delta} \sum_{t,
      \mathbf{x}} \eta_{t, \mathbf{x}}^2}.
\end{equation}
To be explicit, in the usual derivation of the MSR functional this average is
written in the following way:
\begin{equation}
  \label{eq:Villain-666}
  \langle O[\theta] \rangle = \int \mathcal{D}[\eta] \mathcal{P}[\eta]
  O[\theta_{\eta}] = \int \mathcal{D}[\theta, \eta] \mathcal{P}[\eta]
  O[\theta] \delta[\theta - \theta_{\eta}],
\end{equation}
where
\begin{equation}
  \label{eq:Villain-65}
  \begin{split}
    \int \mathcal{D}[\eta] & = \prod_{t, \mathbf{x}} \int_{-\infty}^{\infty} d
    \eta_{t, \mathbf{x}}, \\ \int \mathcal{D}[\theta] & = \prod_{t, \mathbf{x}}
    \int_0^{2 \pi} d \theta_{t, \mathbf{x}}.
  \end{split}
\end{equation}
In the second equality in Eq.~\eqref{eq:Villain-666}, we introduced an
additional integration over $\theta$, which is fixed to $\theta_{\eta}$ by the
$\delta$-functional. The latter can be expressed as the product over
$\delta$-functions for the values of $\theta_{t, \mathbf{x}}$ at specific points
$(t, \mathbf{x})$ on the space-time lattice:
\begin{equation}
  \label{eq:Villain-5}
  \begin{split}
    \delta[\theta - \theta_{\eta}] & = \prod_{t, \mathbf{x}} \delta(\theta_{t,
      \mathbf{x}} - \theta_{\eta, t, \mathbf{x}}) \\ & = \prod_{t, \mathbf{x}}
    \sum_{n_{t, \mathbf{x}}} \delta(\Delta_t \theta_{t, \mathbf{x}} - \epsilon
    \left( \mathcal{L}[\theta]_{t, \mathbf{x}} + \eta_{t, \mathbf{x}} \right) +
    2 \pi n_{t, \mathbf{x}}).
  \end{split}
\end{equation}
In the last equality, we used that the Jacobian of the operator
$\Delta_t \theta_{t, \mathbf{x}} - \epsilon \mathcal{L}[\theta]_{t, \mathbf{x}}$
for the retarded regularization chosen in Eq.~\eqref{eq:Villain-78} is equal to
one. As pointed out above, for a given value of $\theta_{t, \mathbf{x}}$, the
value of $n_{t, \mathbf{x}}$ is uniquely specified. In other words, for a given
value of $\theta_{t, \mathbf{x}}$ there is a unique combination of
$\theta_{t + \epsilon, \mathbf{x}} \in [0, 2\pi)$ and $n_{t, \mathbf{x}} \in \Z$
for which the argument of the $\delta$-function becomes zero. Hence, taking the
sum over $n_{t, \mathbf{x}} \in \Z$ in Eq.~\eqref{eq:Villain-5} does not
correspond to an additional averaging.

Evaluating Eq.~\eqref{eq:Villain-5} further, instead of with a single
$\delta$-functional as in the usual MSR construction, we have to deal with a
train of $\delta$-functions, which can in turn be rewritten as a Fourier sum
instead of a Fourier integral as in the usual
case~\cite{Kamenev2011,Altland2010a,Tauber2014a}:
\begin{multline}
  \label{eq:Villain-8}
  \sum_{n_{t, \mathbf{x}} = -\infty}^{\infty} \delta(\Delta_t \theta_{t,
    \mathbf{x}} - \epsilon \left( \mathcal{L}[\theta]_{t, \mathbf{x}} + \eta_{t,
      \mathbf{x}} \right) + 2 \pi n_{t, \mathbf{x}}) \\ = \frac{1}{2 \pi}
  \sum_{\tilde{n}_{t, \mathbf{x}} = -\infty}^{\infty} e^{-i \tilde{n}_{t,
      \mathbf{x}} \left[ \Delta_t \theta_{t, \mathbf{x}} - \epsilon \left(
        \mathcal{L}[\theta]_{t, \mathbf{x}} + \eta_{t, \mathbf{x}} \right)
    \right]}.
\end{multline}
In consequence, the ``response field'' $\tilde{n}$ is integer-valued and not
continuous as in the non-compact case. The next step in the construction of the
MSR functional integral is to insert Eq.~\eqref{eq:Villain-8} in
Eq.~\eqref{eq:Villain-5} and perform the integration over the noise field $\eta$
which leads to
\begin{equation}
  \label{eq:Villain-9}  
  \langle O[\theta] \rangle \propto \sum_{\{ \tilde{n}_{t, \mathbf{x}} \}}
  \int \mathcal{D}[\theta] O[\theta] e^{i S}.
\end{equation}
The exponent in this expression defines the MSR action,
\begin{equation}
  \label{eq:Villain-11}  
  S = \sum_{t, \mathbf{x}} \tilde{n}_{t, \mathbf{x}} \left[ - \Delta_t
  \theta_{t, \mathbf{x}} + \epsilon \left( \mathcal{L}[\theta]_{t, \mathbf{x}} + i
   \Delta \tilde{n}_{t, \mathbf{x}} \right) \right],
\end{equation}
and the MSR functional integral is thus given by
\begin{equation}
  \label{eq:Villain-10}
  Z = \sum_{\{ \tilde{n}_{t, \mathbf{x}} \}} \int \mathcal{D}[\theta] e^{i S}.
\end{equation}
Note that because $\tilde{n}_{t, \mathbf{x}}$ is integer-valued, the weight
$e^{i S}$ in the MSR functional integral is indeed invariant under the
above-mentioned gauge transformation
$\theta_{t, \mathbf{x}} \mapsto \theta_{t, \mathbf{x}} + 2 \pi n_{t,
  \mathbf{x}}$
as it should be. This is formally similar to the Matsubara functional integral
description of Josephson junction array
models~\cite{Fisher1989a,Fazio1991,Fazio2001}, where $\tilde{n}$ corresponds to
the charge, and the invariance under shifts of the phase by multiples of $2 \pi$
is guaranteed by the discreteness of the latter.

Finally, we rewrite the action in a form that is more convenient for performing
the duality transformation below. It is straightforward to verify the equality
\begin{multline}
  \label{eq:Villain-67}  
  \epsilon \sum_{t, \mathbf{x}} \tilde{n}_{t, \mathbf{x}} \mathcal{L}[\theta]_{t,
    \mathbf{x}} \\ = - \epsilon \sum_{t, \mathbf{x}, i} \left[ D \left(
      \tilde{n}_{t, \mathbf{x}} - \tilde{n}_{t, \mathbf{x} + \unitvec{e}_i} \right)
    \sin(\theta_{t, \mathbf{x}} - \theta_{t, \mathbf{x} + \unitvec{e}_i})
    \vphantom{\frac{\lambda}{2}} \right. \\ \left. + \frac{\lambda}{2} \left(
    \tilde{n}_{t, \mathbf{x}} + \tilde{n}_{t, \mathbf{x} + \unitvec{e}_i} \right)
  \left( \cos(\theta_{t, \mathbf{x}} - \theta_{t, \mathbf{x} + \unitvec{e}_i}) -
    1 \right) \right],
\end{multline}
where $i = x,y$, and $\unitvec{e}_i$ denote the unit vectors in the respective
directions. Ultimately, we are interested in the continuum limit both with
respect to time and space. Therefore, in the term
$\left( \tilde{n}_{t, \mathbf{x}} + \tilde{n}_{t, \mathbf{x} + \unitvec{e}_i}
\right) \cos(\theta_{t, \mathbf{x}} - \theta_{t, \mathbf{x} + \unitvec{e}_i})$
we can replace $\tilde{n}_{t, \mathbf{x} + \unitvec{e}_i}$ by
$\tilde{n}_{t, \mathbf{x}}$, since the difference vanishes in this limit. Then,
to leading order in $\epsilon$, we have the following equality (to make the
notation more compact, in the following we denote the lattice derivative by
$\Delta_i \theta_{t, \mathbf{x}} = \theta_{t, \mathbf{x} + \mathbf{e}_i} -
\theta_{t, \mathbf{x}}$):
\begin{multline}
  \label{eq:Villain-686}
  \epsilon \sum_{t, \mathbf{x}} \tilde{n}_{t, \mathbf{x}} \mathcal{L}[\theta]_{t,
    \mathbf{x}} = - \sum_{t, \mathbf{x}, i} \left[ \sin(\Delta_i \theta_{t,
      \mathbf{x}}) \sin(\epsilon D \Delta_i \tilde{n}_{t, \mathbf{x}}) \right. \\
  \left. + \epsilon \lambda \tilde{n}_{t, \mathbf{x}} \left( \cos(\Delta_i
      \theta_{t, \mathbf{x}}) \cos(\epsilon D \Delta_i \tilde{n}_{t, \mathbf{x}})
      - 1 \right) \right] + O(\epsilon^3),
\end{multline}
which can be seen to reduce to Eq.~\eqref{eq:Villain-67} straightforwardly by
expanding the trigonometric functions in powers of $\epsilon$. As a last step, we
introduce an additional summation over $\sigma = \pm 1$ which allows us to write
the RHS of Eq.~\eqref{eq:Villain-686} in the form
\begin{multline}
  \label{eq:Villain-68}
  \epsilon \sum_{t, \mathbf{x}} \tilde{n}_{t, \mathbf{x}} \mathcal{L}[\theta]_{t,
    \mathbf{x}} = \frac{1}{2} \sum_{t, \mathbf{x}, i, \sigma} \left( \sigma -
    \epsilon \lambda \tilde{n}_{t, \mathbf{x}} \right) \\ \times \left(
    \cos(\Delta_i (\theta_{t, \mathbf{x}} + \sigma \epsilon D \tilde{n}_{t,
      \mathbf{x}})) - 1 \right) + O(\epsilon^3).
\end{multline}
Omitting higher order corrections in $\epsilon$, and using a compact notation in
which $X = \left( t, \mathbf{x} \right)$, the MSR action~\eqref{eq:Villain-11}
can be written as
\begin{multline}
  \label{eq:Villain-13}
  S = \sum_X \biggl[ \tilde{n}_X \left( - \Delta_t \theta_X + i \epsilon \Delta
    \tilde{n}_X \right) \\ + \frac{1}{2} \sum_{i, \sigma} \left( \sigma - \epsilon
    \lambda \tilde{n}_X \right) \left( \cos(\Delta_i (\theta_X + \sigma \epsilon D
    \tilde{n}_X)) - 1 \right) \biggr].
\end{multline}
Note that the action obeys causality in the sense of Keldysh field
theory~\cite{Kamenev2011,Altland2010a}, i.e., $S = 0$ for $\tilde{n} = 0$, which
is an essential property of Keldysh and MSR functional
integrals~\cite{Kamenev2011,Altland2010a,Tauber2014a}. For the dynamical and
noise terms (the first line on the RHS of Eq.~\eqref{eq:Villain-13}), as well as
for the term that is proportional to $\lambda$, this is evident: these terms are
of linear or higher order in $\tilde{n}$; the remaining part of the action
resembles a typical Hamiltonian contribution to a Keldysh action, i.e., a term
of the form $\sum_{\sigma} \sigma \mathcal{H}[\theta_{\sigma}]$. In the present
case,
\begin{equation}
  \label{eq:Villain-69}
  \mathcal{H}[\theta_{\sigma}] = - \frac{1}{2} \sum_{X, i} \left(
    \cos(\Delta_i \theta_{\sigma, X}) - 1 \right),
\end{equation}
where $\theta_{\sigma, X} = \theta_X + \sigma \epsilon D \tilde{n}_X$ are to some
extent analogous to fields on the forward and backward branches of the Keldysh
contour. Then, for $\tilde{n} = 0,$ it follows that $\theta_+ = \theta_-$, and
the Hamiltonian part of the action vanishes due to the summation over $\sigma$.

\subsection{Duality transformation}
\label{sec:dual-transf}

The form of the action~\eqref{eq:Villain-13} is the starting point for
performing the duality transformation. To this end we write the MSR partition
function~\eqref{eq:Villain-10} in the form
\begin{multline}
  \label{eq:Villain-15}  
  Z = \sum_{\{ \tilde{n}_X \}} \int \mathcal{D}[\theta] e^{i \sum_X \tilde{n}_X
    \left( - \Delta_t \theta_X + i \epsilon \Delta \tilde{n}_X \right)} \\ \times
  \prod_{X, i, \sigma} e^{i K_{\sigma X} \left( \cos(\Delta_i (\theta_X +
      \sigma \epsilon D \tilde{n}_X)) - 1 \right)},
\end{multline}
where the prefactor in the exponent is
\begin{equation}
  \label{eq:Villain-16}
  K_{\sigma X} = \frac{1}{2} \left( \sigma - \epsilon \lambda \tilde{n}_X
  \right).
\end{equation}
The exponential in the second line in Eq.~\eqref{eq:Villain-15} is a periodic
function of the variable $\Delta_i (\theta_X + \sigma \epsilon D \tilde{n}_X)$
which is defined in terms of the values of $\theta$ and $\tilde{n}$ on the two
sites $\left( t, \mathbf{x} \right)$ and
$\left( t, \mathbf{x} + \unitvec{e}_i \right)$ of the spatio-temporal
lattice. Hence, we can expand this exponential as a Fourier series in which the
sum is taken over a new variable $j_{\sigma i X}$ associated with the link
connecting these lattice sites,\footnote{According to Eq.~\eqref{eq:Villain-8}
  also $\tilde{n}$ is defined on links connecting sites
  $\left( t, \mathbf{x} \right)$ and $\left( t + \epsilon, \mathbf{x} \right)$
  which are separated in the direction of time. This can be seen by noting that
  both $\Delta_t \theta_{t, \mathbf{x}}$ and the increment
  $\epsilon \left( \mathcal{L}[\theta]_{t, \mathbf{x}} + \eta_{t, \mathbf{x}}
  \right)$ are defined on such links.}
\begin{multline}
  \label{eq:Villain-17}  
  e^{i K_{\sigma X} \left( \cos(\Delta_i (\theta_X + \sigma \epsilon D
      \tilde{n}_X)) - 1 \right)} \\ = \sum_{j_{\sigma i X} = -\infty}^{\infty}
  e^{- i \sigma j_{\sigma i X} \Delta_i (\theta_X + \sigma \epsilon D \tilde{n}_X)
    + i V_{\sigma X}(j_{\sigma i X})}.
\end{multline}
Our choice of the sign $- \sigma$ in the first term in the exponent on the RHS
ensures causality as explained below. The presence of the prefactor
$K_{\sigma X}$ in the exponent on the LHS implies that also the coefficients in
the Fourier series, which we write in the form
$e^{i V_{\sigma X}(j_{\sigma i X})}$, depend on $\sigma$ and $X$, as is
indicated by the subscript in the potential $V_{\sigma X}$. In the Villain
approximation (more details are provided in App.~\ref{sec:vill-appr} below), the
latter assumes the form
\begin{equation}
  \label{eq:Villain-18}
  e^{i V_{\sigma X}(j_{\sigma i X})} \approx \frac{1}{\sqrt{i 2 \pi
      K_{\sigma X}}} e^{i j_{\sigma i X}^2/(2 K_{\sigma X})}.
\end{equation}
This leads to the following contribution in the functional integral in
Eq.~\eqref{eq:Villain-15} (here we use the shorthand
$j_{\sigma X}^2 = \abs{\mathbf{j}_{\sigma X}}^2 = j_{\sigma x X}^2 + j_{\sigma y
  X}^2$):
\begin{multline}
  \label{eq:Villain-19}
  \prod_{X, i, \sigma} e^{i K_{\sigma X} \left( \cos(\Delta_i (\theta_X +
      \sigma \epsilon D \tilde{n}_X)) - 1 \right)} \\ = C \sum_{\{
    \mathbf{j}_{\sigma X} \}} e^{i \sum_{X, \sigma} \left[ - \sigma
      \mathbf{j}_{\sigma X} \cdot \nabla (\theta_X + \sigma \epsilon D \tilde{n}_X)
      + j_{\sigma X}^2/(2 K_{\sigma X}) \right]},
\end{multline}
where in the second line we denote $\nabla = \left( \Delta_x, \Delta_y \right)$.
The prefactor is (in 2D the product over directions gives a square)
\begin{equation}
  \label{eq:Villain-20}
  C = \prod_{X, i, \sigma} \frac{1}{\sqrt{i 2 \pi K_{\sigma X}}} = \prod_{X,
    \sigma} \frac{1}{i 2 \pi K_{\sigma X}}.
\end{equation}
We note that in terms of the newly introduced variables $\mathbf{j}_{\sigma X}$
the action is still causal in the above-mentioned sense, i.e., we have $S = 0$
for $\tilde{n} = 0$ and $\mathbf{j}_{+} = \mathbf{j}_{-}$. To see this, first note
that for $\tilde{n} = 0$ in Eq.~\eqref{eq:Villain-16} we have
$K_{\sigma X} = \sigma/2$. Then, in the sum in the exponent in
Eq.~\eqref{eq:Villain-19} there is an overall prefactor $\sigma$, and therefore
the sum vanishes for $\mathbf{j}_{+ X}^2 = \mathbf{j}_{- X}^2$.

As indicated above, eventually we are interested in the continuum limit in time,
$\epsilon \to 0$. Hence, in the following we keep only the leading terms in
$\epsilon$. This allows us to considerably simplify the factor $C$ defined in
Eq.~\eqref{eq:Villain-20}. Indeed, reexponentiating $K_{\sigma X}$ and expanding
the logarithm to second order we find (in the following we do not keep track of
purely numerical, i.e., field-independent, factors; they are inconsequential for
our considerations and can be absorbed in the integration measure in the MSR
functional integral Eq.~\eqref{eq:Villain-15})
\begin{equation}
  \label{eq:Villain-21}  
  \begin{split}
    C & \propto \prod_{X, \sigma} e^{- \ln K_{\sigma X}} = \prod_{X, \sigma} 2
    \sigma e^{- \sigma \epsilon \lambda \tilde{n}_X - \frac{1}{2} \left( \epsilon
        \lambda \tilde{n}_X \right)^2 + O(\epsilon^3)} \\ & \propto e^{- \left(
        \epsilon \lambda \right)^2 \sum_X \tilde{n}_X^2 + O(\epsilon^3)}.
  \end{split}
\end{equation}
The first order term in the exponent vanishes upon taking the product of the
exponential over $\sigma$ (or, equivalently, the sum over $\sigma$ in the
exponent). As a result, $C$ gives a contribution to the MSR action of second
order in $\epsilon$ and can hence be ignored.

Now we examine the last term in the exponent in Eq.~\eqref{eq:Villain-19}. Expanding
$1/K_{\sigma X}$ to first order in $\epsilon$ we find
\begin{equation}
  \label{eq:Villain-22}
  \frac{i}{2} \sum_{X, \sigma} \frac{j_{\sigma X}^2}{K_{\sigma X}} = i
  \sum_X \left[ \mathbf{j}_X \cdot \tilde{\bj}_X + \frac{\epsilon \lambda}{2}
    \tilde{n}_X \left( j_X^2 + \tilde{j}_X^2 \right) \right],
\end{equation}
where we introduce the current $\mathbf{j} = \mathbf{j}_{+} + \mathbf{j}_{-}$
and the response current $\tilde{\bj} = \mathbf{j}_{+} - \mathbf{j}_{-}$. Below
in Eq.~\eqref{eq:Villain-36} we replace the integer-valued vector fields
$\mathbf{j}$ and $\tilde{\bj}$ by continuous ones by means of the Poisson
summation formula. This is a necessary prerequisite for taking the limit
$\epsilon \to 0$, for the simple reason that a sequence of integers at times
$t, t + \epsilon, t + 2 \epsilon, \dotsc$ cannot converge to a continuous
function of time for $\epsilon \to 0$. Moreover, taking a sensible
$\epsilon \to 0$ limit requires us to rescale the real-valued response current
as $\tilde{\bj} \to \epsilon \tilde{\bj}$, as becomes clear from
Eq.~\eqref{eq:Villain-24} below. Anticipating these steps, we see that the
contribution involving $\tilde{j}_X^2$ in Eq.~\eqref{eq:Villain-22} above can
actually be considered to be $O(\epsilon^3)$, and we discard it already at this
point.

Putting the pieces together, the partition function~\eqref{eq:Villain-15}
becomes
\begin{multline}
  \label{eq:Villain-23}  
  Z \propto \sum_{\{\tilde{n}_X, \mathbf{j}_X, \tilde{\bj}_X \}} \int
  \mathcal{D}[\theta] e^{i \sum_X \theta_X \left( \Delta_t \tilde{n}_X + \nabla
      \cdot \tilde{\bj}_X \right)} \\ \times e^{i \sum_X \left[ \epsilon \tilde{n}_X
      \left( D \nabla \cdot \mathbf{j}_X + i \Delta \tilde{n}_X \right) +
      \mathbf{j}_X \cdot \left( \tilde{\bj}_X + \frac{\epsilon \lambda}{2} \tilde{n}_X
        \mathbf{j}_X \right) \right]}.
\end{multline}
In the exponent we summed by parts twice. We note that the exponent is linear
both in $\theta$ and $\tilde{\bj}$. Hence, the sum over $\tilde{\bj}$ can be
carried out and gives
$\prod_X \sum_{\mathbf{m}_X} \delta(- \nabla \theta_X + \mathbf{j}_X - 2 \pi
\mathbf{m}_X)$
with an integer-valued vector field $\mathbf{m}$. Some intuition can be gained
by decomposing this vector field into longitudinal and transverse parts. The
longitudinal part can be written as the lattice gradient of an integer field
$m_l$, and the transverse part as the lattice curl of a vector field
$m_t \unitvec{z}$ pointing along the $z$-direction. Absorbing the longitudinal
component $m_l$ into $\theta$ by extending the integration in
Eq.~\eqref{eq:Villain-65} over the whole real axis (in other words, making
$\theta$ non-compact) the argument of the $\delta$-function suggests that (up to
a prefactor) we can interpret $\mathbf{j}$ as the bosonic current. The latter
has both a smooth longitudinal contribution $\nabla \theta$ and --- in the
presence of vortices --- a transverse component corresponding to non-zero values
of $m_t$.

However, instead of summing over $\tilde{\bj}$ in Eq.~\eqref{eq:Villain-23}, we
take a different route and integrate out $\theta$. This yields a
$\delta$-function corresponding to the constraint that $\tilde{n}_X$ and
$\tilde{\bj}_X$ should satisfy the continuity equation,
\begin{equation}
  \label{eq:Villain-24}
  \Delta_t \tilde{n}_X + \nabla \cdot \tilde{\bj}_X = 0.
\end{equation}
(Note that as indicated above this equation implies that the continuous fields
which replace $\tilde{n}$ and $\tilde{\bj}$ scale differently in the limit
$\epsilon \to 0$.) Formally, the appearance of a continuity equation is again
analogous to the duality transformation in a quantum system at
$T = 0$~\cite{Fisher1989a,Fazio1991,Fazio2001}. However, in the latter case, the
continuity equation is a consequence of particle number conservation. In
driven-dissipative condensates, on the other hand, the number of particles is
\emph{not} conserved. Nevertheless, there is a residual $U(1)$ phase-rotation
symmetry~\cite{Sieberer2015a} which is reflected in the appearance of a
Goldstone boson in the (mean-field) condensed phase (indeed, the KPZ
equation~\eqref{eq:KPZ} is the massless equation of motion of the Goldstone
boson which is the phase of the condensate), and in the above continuity
equation for the \emph{response} fields $\tilde{n}$ and $\tilde{\bj}$. From this
continuity equation, the Noether charge associated with the $U(1)$ symmetry can
be seen to be the sum over space of $\tilde{n}_X$. However, by construction of
the MSR formalism, the response fields have vanishing expectation value, and
therefore the Noether charge is always zero and the continuity
equation~\eqref{eq:Villain-24} is trivially satisfied on average.

The continuity equation~\eqref{eq:Villain-24} can be interpreted as stating that
the three-component vector field $(\tilde{n}, \tilde{\bj})$ has vanishing
divergence. Hence, this vector field can be parametrized as the curl of another
vector field,
\begin{equation}
  \label{eq:Villain-25}
  \begin{pmatrix}
    \tilde{n} \\ \tilde{\bj}
  \end{pmatrix} =
  \begin{pmatrix}
    \Delta_t \\ \nabla
  \end{pmatrix} \times
  \begin{pmatrix}
    \tilde{\phi} \\ - \tilde{\mathbf{A}}
  \end{pmatrix} =
  \begin{pmatrix}
    - \unitvec{z} \cdot \left( \nabla \times \tilde{\mathbf{A}} \right) \\
    - \unitvec{z} \times \left( \nabla \tilde{\phi} + \Delta_t
      \tilde{\mathbf{A}} \right)
  \end{pmatrix}
  .
\end{equation}
Here and in the following it is understood, that --- depending on the context
--- the gradient operator and vectors such as $\tilde{\mathbf{A}}$ should be
considered as having two components or three components with the third one being
zero. The parametrization of $(\tilde{n}, \tilde{\bj})$ in terms of the
potentials $(\tilde{\phi}, \tilde{\mathbf{A}})$ is not unique. In fact, the
``physical'' fields $(\tilde{n}, \tilde{\bj})$ are invariant under the gauge
transformation
\begin{equation}
  \label{eq:Villain-26}
  \begin{split}
    \tilde{\phi} & \to \tilde{\phi} - \Delta_t \chi, \\
    \tilde{\mathbf{A}} & \to \tilde{\mathbf{A}} + \nabla \chi,
  \end{split}
\end{equation}
with an arbitrary integer field $\chi$. We can exploit this freedom by choosing
a gauge that leads to a simple form of the action. However, for the moment we
leave the gauge unspecified. Then, summing over both $\tilde{\phi}$ and
$\tilde{\mathbf{A}}$ without restriction simply introduces a multiplicative
overcounting in the partition function~\eqref{eq:Villain-23}.

In the following, we find it instructive to parametrize also the current $\bj$
in terms of gauge potentials --- although, strictly speaking, this is not
necessary. The parametrization is chosen in analogy to Eq.~\eqref{eq:Villain-25}
for the response current, however, omitting the time derivative of the vector
potential. Hence, we set
\begin{equation}
  \label{eq:Villain-27}
  \mathbf{j} = - \unitvec{z} \times \left( \nabla \phi + \mathbf{A} \right).
\end{equation}
As above, this parametrization is not unique. Different, physically equivalent
possibilities are related by gauge transformations, which read in the present
case
\begin{equation}
  \label{eq:Villain-28}
  \begin{split}
    \phi & \to \phi - \chi, \\
    \mathbf{A} & \to \mathbf{A} + \nabla \chi.
  \end{split}
\end{equation}
The quantities $\phi$ and $\mathbf{A}$ are the scalar and vector potentials of
the dissipative electrodynamics introduced heuristically in Ref.~\cite{Wachtel2016} and described
below in Sec.~\ref{sec:dual-electr-theory}. In terms of these potentials, the
gauge-invariant electric and magnetic fields are defined as
\begin{equation}
  \label{eq:Villain-30}
  \begin{split}
    \mathbf{E} & = - \unitvec{z} \times \mathbf{j} = - \nabla \phi - \mathbf{A},
    \\
    \mathbf{B} & = D \nabla \times \mathbf{A}.
  \end{split}
\end{equation}
Note that since both $\nabla$ and $\mathbf{A}$ have \emph{vanishing} components
in the $z$-direction, the \emph{only non-vanishing} component of the magnetic
field $\mathbf{B}$ is exactly along $\unitvec{z}$, i.e.,
$\mathbf{B} = B \unitvec{z}$. This implies that the homogeneous Maxwell equation
\begin{equation}
  \label{eq:divB}
  \nabla \cdot \mathbf{B} = 0,  
\end{equation}
is trivially satisfied. Moreover, due to the absence of the usual time
derivative acting on the vector potential $\mathbf{A}$ in the definition of the
electric field $\mathbf{E}$ in Eq.~\eqref{eq:Villain-30}, the magnetic field can
be expressed directly in terms of the electric field by means of a modified
Faraday's law~\cite{Wachtel2016},
\begin{equation}
  \label{eq:Faraday}
  \nabla \times \mathbf{E} + \frac{1}{D} \mathbf{B} = 0.
\end{equation}
According to the expression of the electric field in terms of the current in
Eq.~\eqref{eq:Villain-30}, we find
$\nabla \times \mathbf{E} = - \left( \nabla \cdot \mathbf{j} \right)
\unitvec{z}$.
Inserting this relation in Eq.~\eqref{eq:Faraday} and keeping in mind that
$\mathbf{B} = B \unitvec{z}$, we see that Faraday's law is just the continuity
equation, where the magnetic field encodes fluctuations of the bosonic density
$\rho$ around the mean value $\rho_0$, i.e.,
$B \propto - \left( \rho - \rho_0 \right)$. The same identification is made in
the dual electrodynamic theory for superfluid films in thermal
equilibrium~\cite{Ambegaokar1978,Ambegaokar1980}. A crucial difference is that
in a driven-dissipative system without particle number conservation the
continuity equation includes a source term
$\propto \left( \rho - \rho_0 \right)$~\cite{Altman2015} which dominates over
the usual term $\partial_t \rho$ in the low-frequency limit. It is precisely
this limit (in which fluctuations of the phase of a driven-dissipative
condensate are described by the KPZ equation) which we are considering here.

Inserting Eqs.~\eqref{eq:Villain-25},~\eqref{eq:Villain-27}, and
\eqref{eq:Villain-30} in the partition function in Eq.~\eqref{eq:Villain-23},
the latter becomes
\begin{equation}
  \label{eq:Villain-29}
  Z \propto \sum_{\{ \phi_X, \tilde{\phi}_X, \mathbf{A}_X, \tilde{\mathbf{A}}_X
    \}} e^{i S_{\mathit{EB}}},
\end{equation}
where the action is given by
\begin{multline}
  \label{eq:Villain-31}
  S_{\mathit{EB}} = \sum_X \left[ \tilde{\phi}_X \nabla \cdot \mathbf{E}_X +
    \tilde{\mathbf{A}}_X \cdot \left( \Delta_t
      \mathbf{E}_X - \epsilon \nabla \times \mathbf{B}_X
      \vphantom{\frac{\lambda}{2}} \right. \right. \\
  \left. \left. + \frac{\epsilon \lambda}{2} \unitvec{z} \times \nabla E_X^2
    \right) + i \epsilon \Delta \left( \nabla \times \tilde{\mathbf{A}}_X
    \right)^2 \right].
\end{multline}

For completeness we mention that in analogy to Eq.~\eqref{eq:Villain-30}, we
could define response electric and magnetic fields, which are invariant under
the gauge transformation given in Eq.~\eqref{eq:Villain-26}, as
\begin{equation}
  \label{eq:Villain-32}
  \begin{split}
    \tilde{\mathbf{E}} & = - \nabla \tilde{\phi} - \Delta_t \tilde{\mathbf{A}}, \\
    \tilde{\mathbf{B}} & = \nabla \times \tilde{\mathbf{A}}.
  \end{split}
\end{equation}
Then, the action~\eqref{eq:Villain-31} can be written in a manifestly
gauge-invariant form as
\begin{multline}
  \label{eq:Villain-33}
  S_{\mathit{EB}} = \sum_X \left[ \tilde{\mathbf{E}}_X \cdot \mathbf{E}_X
    \vphantom{\left( \frac{\lambda}{2} \right)} \right. \\ \left. + \epsilon
    \tilde{\mathbf{B}}_X \cdot \left( \mathbf{B}_X + \frac{\lambda}{2} \unitvec{z}
      E_X^2 \right) + i \epsilon \Delta \tilde{B}_X^2 \right].
\end{multline}

In the action in Eq.~\eqref{eq:Villain-31}, the terms multiplying the response
scalar and vector potentials $\tilde{\phi}$ and $\tilde{\mathbf{A}}$ are
reminiscent of the inhomogeneous Maxwell equations, i.e., of Gauss' law and
Amp\`ere's law (enriched by the KPZ non-linearity), however, with the source
terms missing. To fully establish the equivalence to these Maxwell equations, we
introduce charges $n_v, \tilde{n}_v$ and currents
$\mathbf{J}_v, \tilde{\mathbf{J}}_v$ by means of the Poisson summation
formula. The latter reads for a general function $g(k)$ (see,
e.g.,~\cite{Chaikin1995}):
\begin{equation}
  \label{eq:2}
  \sum_{k = -\infty}^{\infty} g(k) = \sum_{n = -\infty}^{\infty}
  \int_{-\infty}^{\infty} d \phi \, g(\phi) e^{-i 2 \pi n \phi}.
\end{equation}
Applying this relation to the summations over $\phi, \tilde{\phi}, \mathbf{A},$
and $\tilde{\mathbf{A}}$ in Eq.~\eqref{eq:Villain-29}, we obtain
\begin{equation}
  \label{eq:Villain-36}
  Z \propto \sum_{\substack{ \{ n_{v X}, \tilde{n}_{v X}, \\
    \mathbf{J}_{v X}, \tilde{\mathbf{J}}_{v X} \}}} \int \mathcal{D}[\phi, \tilde{\phi},
  \mathbf{A}, \tilde{\mathbf{A}}] e^{i S[\phi, \tilde{\phi}, \mathbf{A},
    \tilde{\mathbf{A}}, n_v, \tilde{n}_v, \mathbf{J}_v, \tilde{\mathbf{J}}_v]}.
\end{equation}
Here, we have already included the new summation variables in the action, which
reads
\begin{multline}
  \label{eq:Villain-35}
  S = \sum_X \left[ \tilde{\phi}_X \left( \nabla \cdot \mathbf{E}_X - 2 \pi n_{v
        X} \right) + \tilde{\mathbf{A}}_X \cdot \left( \Delta_t \mathbf{E}_X
      \vphantom{\frac{\lambda}{2}} \right. \right. \\ \left. \left. - \epsilon
      \nabla \times \mathbf{B}_X + 2 \pi \mathbf{J}_{v X} + \frac{\epsilon
        \lambda}{2} \unitvec{z} \times \nabla E_X^2 \right) \right. \\
  \left. \vphantom{\left( \frac{\lambda}{2} \right)} + i \epsilon \Delta \left(
      \nabla \times \tilde{\mathbf{A}}_X \right)^2 - 2 \pi \left( \tilde{n}_{v
        X} \phi_X - \tilde{\mathbf{J}}_{v X} \cdot \mathbf{A}_X \right) \right].
\end{multline}

The Poisson summation formula allowed us to replace the summations in
Eq.~\eqref{eq:Villain-29} over integer-valued fields by integrals over
corresponding continuous fields, at the expense of introducing additional
summation variables. This, however, is a price we are willing to pay: as
indicated above and as we show in the following, the new variables have a clear
physical interpretation as (vortex) charge and current densities, acting as
sources for the electric and magnetic fields. To establish this identification,
we have to make the action Eq.~\eqref{eq:Villain-35} gauge invariant. Indeed,
before using the Poisson summation formula in Eq.~\eqref{eq:Villain-36}, the
action was fully gauge-invariant (cf.\ Eq.~\eqref{eq:Villain-33}). However, the
additional terms appearing in the action (those involving the vortex charge
densities and current densities, $n_v, \tilde{n}_v$ and
$\mathbf{J}_v, \tilde{\mathbf{J}}_v$, respectively, in
Eq.~\eqref{eq:Villain-35}) after replacing sums by integrals according to
Eq.~\eqref{eq:2} break gauge invariance. The latter can be restored by means of
the following trick: in the partition function~\eqref{eq:Villain-36}, we shift
the integration variables according to the gauge transformation prescriptions
given in Eqs.~\eqref{eq:Villain-26} and~\eqref{eq:Villain-28}. Then, in the
action the fields $\chi$ and $\tilde{\chi}$ drop out in all terms with the
exception of those which are not gauge-invariant, i.e, the ones involving the
charges and currents. To be precise, under the gauge transformation the action
becomes
\begin{multline}
  \label{eq:Villain-37}    
  S[\phi - \chi, \tilde{\phi} - \Delta_t \tilde{\chi}, \mathbf{A} + \nabla \chi,
  \tilde{\mathbf{A}} + \nabla \tilde{\chi}, n_v, \tilde{n}_v, \mathbf{J}_v,
  \tilde{\mathbf{J}}_v] \\ = S[\phi, \tilde{\phi}, \mathbf{A},
  \tilde{\mathbf{A}}, n_v, \tilde{n}_v, \mathbf{J}_v, \tilde{\mathbf{J}}_v] \\
  \,\,\,\,\,\,\,\,\,\,\,\,\, -
  2 \pi \sum_X \left[ \tilde{\chi}_X
    \left( \Delta_t n_{v X} + \nabla \cdot \mathbf{J}_{v X} \right) \right. \\
  \left. - \chi_X \left( \tilde{n}_{v X} - \nabla \cdot \tilde{\mathbf{J}}_{v X}
    \right) \right].
\end{multline}
Since we introduced the fields $\chi, \tilde{\chi}$ in a transformation of
integration variables, the value of the functional integral does not depend on
them (even though they appear explicitly in the action). Hence, performing an
\emph{additional} integration over these fields leads only to an irrelevant
prefactor of the partition function. The benefit of carrying out this
integration is that it allows us to arrange the functional integral as
\begin{equation}
  \label{eq:Villain-38}
  Z \propto \sum_{\substack{\{ n_{v X}, \tilde{n}_{v X}, \\  \mathbf{J}_{v X}, \tilde{\mathbf{J}}_{v X} \}}} \int
  \mathcal{D}[\phi, 
  \tilde{\phi}, \mathbf{A}, \tilde{\mathbf{A}}] e^{i S'[\phi, \tilde{\phi}, \mathbf{A},
    \tilde{\mathbf{A}}, n_v, \tilde{n}_v, \mathbf{J}_v, \tilde{\mathbf{J}}_v]}, 
\end{equation}
with a gauge-invariant action $S'$ that is defined as
\begin{multline}
  \label{eq:Villain-39}
  e^{i S'[\phi, \tilde{\phi}, \mathbf{A}, \tilde{\mathbf{A}}, n_v, \tilde{n}_v,
    \mathbf{J}_v, \tilde{\mathbf{J}}_v]} \\
  \begin{aligned}[t]
    & = \int \mathcal{D}[\chi, \tilde{\chi}] e^{i S[\phi - \chi, \tilde{\phi} +
      \Delta_t \tilde{\chi}, \mathbf{A} + \nabla \chi, \tilde{\mathbf{A}} +
      \nabla \tilde{\chi}, n_v, \tilde{n}_v, \mathbf{J}_v,
      \tilde{\mathbf{J}}_v]} \\ & = \delta[\Delta_t n_v + \nabla \cdot
    \mathbf{J}] \delta[\tilde{n}_v - \nabla \cdot \tilde{\mathbf{J}}_v] e^{i
      S[\phi, \tilde{\phi}, \mathbf{A}, \tilde{\mathbf{A}}, n_v, \tilde{n}_v,
      \mathbf{J}_v, \tilde{\mathbf{J}}_v]}.
  \end{aligned}
\end{multline}
The representation of $S'$ after the first equality shows that the action is now
manifestly gauge-invariant: any further gauge transformation of the fields
appearing in $S'$ can simply be absorbed in a shift of variables in the
integration over $\chi$ and $\tilde{\chi}$. Since the transformed action in
Eq.~\eqref{eq:Villain-37} is linear in $\chi$ and $\tilde{\chi}$, in the second
equality we were able to perform the integrals over these fields explicitly,
which yields two $\delta$-functionals. The first of these expresses conservation
of the number of \emph{vortices}, and in particular, it is not at odds with the
discussion below Eq.~\eqref{eq:Villain-24} concerning the absence of
conservation of the number of \emph{bosons} in a driven-dissipative
condensate. The second $\delta$-functional can be used to evaluate the sum over
$\tilde{n}_v$ in Eq.~\eqref{eq:Villain-38}, whereby $\tilde{n}_v$ is replaced by
$\nabla \cdot \tilde{\mathbf{J}}_v$.

In order to connect the MSR functional integral~\eqref{eq:Villain-38} to the
electrodynamics of Ref.~\cite{Wachtel2016}, which is formulated in terms of
Langevin equations for the electromagnetic fields and charges, we follow the
usual approach~\cite{Kamenev2011,Altland2010a,Tauber2014a} and decouple the
noise vertex in the action by means of a Hubbard-Stratonovich (HS)
transformation,
\begin{equation}
  \label{eq:Villain-41}
  e^{-\epsilon \Delta \sum_X \left( \nabla \times \tilde{\mathbf{A}}_X \right)^2}
  \propto \int \mathcal{D}[\eta] e^{- \epsilon \sum_X \left( \frac{1}{4 \Delta}
      \eta_X^2 + i \tilde{\mathbf{A}}_X \cdot \unitvec{z} \times \nabla \eta_X
    \right)}.
\end{equation}
Then, the partition function becomes
\begin{equation}
  \label{eq:Villain-42}
  Z \propto \!\!\!\! \sum_{\{ n_{v X},
    \mathbf{J}_{v X}, \tilde{\mathbf{J}}_{v X} \}} \int \mathcal{D}[\phi, \tilde{\phi},
  \mathbf{A}, \tilde{\mathbf{A}}, \eta] \delta(\Delta_t n_v + \nabla \cdot
  \mathbf{J}_v) e^{i S}.
\end{equation}
Note that in this form the scalar and vector
potentials appear only implicitly in the electric and magnetic fields, i.e., the
gauge invariance of the action under the gauge transformation of these fields in
Eq.~\eqref{eq:Villain-28} is manifest; it is straightforward to check that the
action is also invariant under gauge transformations~\eqref{eq:Villain-26} of
the response fields. The action in Eq.~\eqref{eq:Villain-42} now reads
\begin{multline}
  \label{eq:Villain-43}
  S = \sum_X \left\{ \tilde{\phi}_X \left( \nabla \cdot \mathbf{E}_X - 2 \pi n_{v X}
    \right) + \tilde{\mathbf{A}}_X \cdot \left[ \Delta_t \mathbf{E}_X
      \vphantom{\frac{\lambda}{2}} \right. \right. \\ \left. - \epsilon \nabla \times
  \mathbf{B}_X + 2 \pi \mathbf{J}_{v X} + \epsilon \unitvec{z} \times \nabla \left(
    \frac{\lambda}{2} E_X^2 + \eta_X \right) \right] \\
\left. \phantom{\left( \frac{\lambda}{2} \right)} - 2 \pi \tilde{\mathbf{J}}_{v X}
  \cdot \mathbf{E}_X + i \frac{\epsilon}{4 \Delta} \eta_X^2 \right\}.
\end{multline}
Due to HS decoupling of the noise vertex, the action is now linear in the
response scalar and vector potentials, $\tilde{\phi}$ and $\tilde{\mathbf{A}}$,
and integration over these fields yields additional $\delta$-functionals,
\begin{multline}
  \label{eq:Villain-44}
  Z \propto \sum_{\{ n_{v X}, \mathbf{J}_{v X}, \tilde{\mathbf{J}}_{v X} \}}
  \int \mathcal{D}[\phi, \mathbf{A}, \eta] \delta[\nabla \cdot \mathbf{E} - 2
  \pi n_v] \\ \times \delta \biggl[ \Delta_t \mathbf{E} - \epsilon \nabla \times
  \mathbf{B} + 2 \pi \mathbf{J}_v + \epsilon \unitvec{z} \times \nabla \left(
    \frac{\lambda}{2} E^2 + \eta \right) \biggr] \\ \times \delta[\Delta_t n_v +
  \nabla \cdot \mathbf{J}_v] e^{- \sum_X \left( i 2 \pi \tilde{\mathbf{J}}_{v X}
      \cdot \mathbf{E}_X + \frac{\epsilon}{4 \Delta} \eta_X^2 \right)}.
\end{multline}
These $\delta$-functionals correspond to the inhomogeneous Maxwell equations of
Ref.~\cite{Wachtel2016}: the first one is Gauss' law (note that in 2D
electrodynamics the usual factor of $4 \pi$ on the RHS is replaced by $2 \pi$),
according to which the charges act as sources for the electric
field. Remembering the relation between the electric field and the current,
Eq.~\eqref{eq:Villain-30}, we see that Gauss' law is just the differential form
of the equation (using continuum notation for clarity)
\begin{equation}
  \label{eq:8}
  \oint_{\partial \Omega} d\mathbf{l} \cdot \mathbf{j} = 2 \pi \int_{\Omega} d
  \mathbf{x} \, n_v, 
\end{equation}
according to which the circulation of the current along the closed boundary
$\partial \Omega$ of an area $\Omega$ is determined by the total vortex charge
within that area. The second $\delta$-functional corresponds to Amp\`ere's law,
which in the present case inherits the non-linearity and the noise source $\eta$
from the KPZ equation~\eqref{eq:KPZ}. Hence, in Eq.~\eqref{eq:Villain-44}, the
interpretation of $n_v$ and $\mathbf{J}_v$ as the vortex and current densities
becomes clear. Actually, the $\delta$-constraint ensuring the continuity
equation for the vortices is redundant: by taking the divergence of Amp\`ere's
law and inserting Gauss' law, it can be seen that the continuity equation is
already contained in the inhomogeneous Maxwell equations.

Equation~\eqref{eq:Villain-44} is already quite close to the dual electrodynamic
theory for driven-dissipative condensates of Ref.~\cite{Wachtel2016}. However,
this equation still assumes discretization in time, and since the vortex and
current densities are integer-valued fields, we cannot take the temporal and
spatial continuum limits in the present form. In the next section, we resolve
this issue by considering a particular representation of $n_v$ and $\mathbf{J}_v$
(Eqs.~\eqref{eq:9} and~\eqref{eq:5} below). Moreover, we give meaning to the term
$\tilde{\mathbf{J}}_v \cdot \mathbf{E}$ appearing in the exponent in the third
line of Eq.~\eqref{eq:Villain-44}.

Before proceeding, we note that as mentioned above Eq.~\eqref{eq:Villain-27}, we
can reach the same result Eq.~\eqref{eq:Villain-44} \emph{without} ever
introducing gauge potentials $\phi$ and $\mathbf{A}$, but rather using
Eqs.~\eqref{eq:Villain-27} and~\eqref{eq:Villain-30} to directly express the
current $\mathbf{j}$ in terms of the electric field $\mathbf{E}$. Then, the MSR
partition function Eq.~\eqref{eq:Villain-29} contains a summation over
$\mathbf{E}$ instead of the double sum over $\phi$ and $\mathbf{A}$. Applying
again the Poisson summation formula as in Eq.~\eqref{eq:Villain-38} leads
directly to a term of the form $\tilde{\mathbf{J}}_v \cdot \mathbf{E}$ that is
also present in Eq.~\eqref{eq:Villain-44}. However, it is instructive to see how
this term emerges from requiring the action to be gauge invariant.

\section{Dual electrodynamic theory}
\label{sec:dual-electr-theory}

Let us consider a collection of vortices at lattice points
$\mathbf{x}_{\alpha}(t)$ where $\alpha = 1, 2, \dotsc$, and with vorticity
$n_{\alpha} \in \Z$. The vortex density $n_v$ corresponding to such a
configuration is given by
\begin{equation}
  \label{eq:9}
  n_{v, t, \mathbf{x}} = \sum_{\alpha} n_{\alpha} \delta_{\mathbf{x},
    \mathbf{x}_{\alpha}(t)}.
\end{equation}
Since the vortex density obeys a continuity equation (expressed by the
$\delta$-functional in the third line of Eq.~\eqref{eq:Villain-44}), the current
density follows immediately from Eq.~\eqref{eq:9} and is given by
\begin{equation}
  \label{eq:5}
  \mathbf{J}_{v, t, \mathbf{x}} = \sum_{\alpha} n_{\alpha} \Delta_t
  \mathbf{x}_{\alpha}(t) \delta_{\mathbf{x}, \mathbf{x}_{\alpha}(t)}.
\end{equation}
By writing the vortex and current densities in this way, they are fully
determined by the vortex trajectories $\mathbf{x}_{\alpha}(t)$. Quite
conveniently, this allows us to take the continuum limit both in space and time:
we simply have to replace the sums over $n_v$ and $\mathbf{J}_v$ in
Eq.~\eqref{eq:Villain-44} by integrals over smooth functions
$\mathbf{x}_{\alpha}(t)$,
\begin{equation}
  \label{eq:10}
  \sum_{\{ n_{v X}, \mathbf{J}_{v X} \}} \to \int \mathcal{D}[\{\mathbf{x}_{\alpha}\}].
\end{equation}
Concomitantly, we replace sums by integrals,
$\sum_{\mathbf{x}} \to \int d \mathbf{x}$ (remember that we set the lattice
spacing to 1) and $\sum_t \epsilon \to \int d t$, discrete derivatives by
ordinary ones, $\Delta_i \to \partial_i$ and $\Delta_t/\epsilon \to \partial_t$,
and finally the Kronecker-$\delta$'s in Eqs.~\eqref{eq:9} and~\eqref{eq:5} by
$\delta$-functions,
$\delta_{\mathbf{x}, \mathbf{x}_{\alpha}(t)} \to \delta(\mathbf{x} -
\mathbf{x}_{\alpha}(t))$.

Before proceeding we comment on a difference between the summation over $n_v$ and
$\mathbf{J}_v$ and the integral over vortex trajectories $\mathbf{x}_{\alpha}$ in
Eq.~\eqref{eq:10}: while in the former case configurations with different
numbers of vortices and antivortices are taken into account, in the latter case
these numbers are fixed by the values of the charges $n_{\alpha}$. Hence, an
additional summation over $\{ n_{\alpha} \}$ should be included. For simplicity,
we consider in the following only a single configuration $\{ n_{\alpha} \}$.

It remains to specify the dynamics of the vortices. Consistent with the
over-damped dynamics of the electric and magnetic fields (cf.\ Faraday's law
Eq.~\eqref{eq:Faraday}), we assume that the vortices undergo diffusive motion.
In the MSR formalism, diffusive motion corresponds to the following contribution
to the action:
\begin{equation}
  \label{eq:4}
  S_d = \frac{1}{\mu} \int dt \sum_{\alpha} \mathbf{p}_{\alpha} \cdot \left( \frac{d
      \mathbf{x}_{\alpha}}{d t} + i T \mathbf{p}_{\alpha} \right).
\end{equation}
This has to be added on phenomenological grounds, as is also the case in the
equilibrium treatment of Ref.~\cite{Ambegaokar1978,Ambegaokar1980}. There, such
a contribution to the action (or, equivalently, to the Langevin equation for the
vortex coordinates $\mathbf{x}_{\alpha}$) ensures that the stationary
distribution is given by a thermal Gibbs ensemble at the vortex ``temperature''
$T$. It is reasonable to assume that the value of the vortex temperature is
close to the dimensionless noise strength $\Delta/D$ in the KPZ
equation~\eqref{eq:KPZ}, since the noise acting on the vortices originates from
the one acting on phase field $\theta$.\footnote{Indeed, in thermal equilibrium
  (i.e., for $\lambda = 0$), the KPZ equation~\eqref{eq:KPZ} reduces to a linear
  diffusion equation describing the relaxation of the phase $\theta$ to a
  thermal stationary distribution at a dimensionless temperature $\Delta/D$,
  which in this case is strictly identical to the vortex temperature.} In
principle, both the vortex temperature $T$ and the vortex mobility $\mu$ could
be determined numerically (see Ref.~\cite{Aranson1998b} for a related discussion
in the context of the complex Ginzburg-Landau equation).

In Eq.~\eqref{eq:4}, $\mathbf{p}_{\alpha}$ is the momentum that is ``conjugate''
(in the sense of the MSR formalism) to the position
$\mathbf{x}_{\alpha}$. Exactly the same relation of mutual conjugacy holds
between the variables $\mathbf{J}_v$ and $\tilde{\mathbf{J}}_v$, suggesting that
similarly to Eq.~\eqref{eq:5}, which expresses $\mathbf{J}_v$ in terms of the
vortex coordinates $\mathbf{x}_{\alpha}$, there should be a representation of
$\tilde{\mathbf{J}}_v$ involving the momenta $\mathbf{p}_{\alpha}$. Indeed, if
we replace $\tilde{\mathbf{J}}_v$ in Eq.~\eqref{eq:Villain-44} according to (at
the same time taking the continuum limit)
\begin{equation}
  \label{eq:3}
  \frac{2 \pi}{\epsilon} \tilde{\mathbf{J}}_{v X} \to \sum_{\alpha} n_{\alpha} \mathbf{p}_{\alpha}(t)
  \delta(\mathbf{x} - \mathbf{x}_{\alpha}(t)),
\end{equation}
and combine the resulting contribution to the action with the one in
Eq.~\eqref{eq:4}, we obtain the complete vortex or charge action
\begin{equation}
  \label{eq:11}
  S_c = \int dt \sum_{\alpha} \mathbf{p}_{\alpha} \cdot \left( \frac{d
      \mathbf{x}_{\alpha}}{d t} - \mu n_{\alpha} \mathbf{E}(\mathbf{x}_{\alpha})
  + i \mu T \mathbf{p}_{\alpha} \right),
\end{equation}
where we additionally rescaled $\mathbf{p}_{\alpha}$ with the vortex mobility
$\mu$. The identification Eq.~\eqref{eq:3} completely removes
$\tilde{\mathbf{J}}_v$ from the action, and correspondingly we replace the
summation $\sum_{\{\tilde{\mathbf{J}}_{v X}\}}$ by an integration over momentum
trajectories $\int \mathcal{D}[\{ \mathbf{p}_{\alpha} \}]$.

Finally, performing a HS decoupling of the noise vertex in the charge action
$S_c$ (cf.\ Eq.~\eqref{eq:Villain-41} above), the latter becomes linear in the
momenta $\mathbf{p}_{\alpha}$. Then, the integration over these variables can be
performed and yields yet another $\delta$-constraint, rendering the functional
integral Eq.~\eqref{eq:Villain-44} in the form
\begin{multline}
  \label{eq:7}
  Z \propto \int \mathcal{D}[\{ \mathbf{x}_{\alpha}, \boldsymbol{\xi}_{\alpha}
  \}, \phi, \mathbf{A}, \eta] \delta[\nabla \cdot \mathbf{E} - 2 \pi n_v] \\
  \times \delta \biggl[ \partial_t \mathbf{E} - \nabla \times \mathbf{B} + 2 \pi
  \mathbf{J}_v + \unitvec{z} \times \nabla \left( \frac{\lambda}{2} E^2 + \eta
  \right) \biggr] \\ \times \prod_{\alpha} \delta \biggl[ \frac{d
    \mathbf{x}_{\alpha}}{d t} - \mu n_{\alpha} \mathbf{E}(\mathbf{x}_{\alpha}) -
  \boldsymbol{\xi}_{\alpha} \biggr] \\ \times e^{- \frac{1}{4 \Delta} \int dt d
    \mathbf{x} \, \eta^2 - \frac{1}{4 T} \int dt \sum_{\alpha}
    \abs{\boldsymbol{\xi}_{\alpha}}^2}.
\end{multline}
By reverting the logic that leads from a Langevin equation to the corresponding
MSR action (cf.\ the discussion in the paragraph above Eq.~(\ref{eq:Villain-7})
in Sec.~\ref{sec:msr-action-compact}) we can see that this functional integral
is equivalent to the electrodynamic theory, which we introduced heuristically in
Ref.~\cite{Wachtel2016}. It is summarized in the set of modified Maxwell
equations, Eqs.~\eqref{eq:divB} and~\eqref{eq:Faraday} (note that these
equations do not appear explicitly in the functional integral Eq.~\eqref{eq:7}
since in the latter the electric and magnetic fields are expressed in terms of
the gauge potentials so that the homogeneous Maxwell equations are satisfied
automatically), Gauss' law
\begin{equation}
  \label{eq:Gauss}
  \nabla \cdot \mathbf{E} = 2 \pi n_v,  
\end{equation}
and Amp\`ere's law,
\begin{equation}
  \label{eq:Ampere}
  \nabla \times \mathbf{B} - \frac{\partial \mathbf{E}}{\partial t} = 2 \pi
  \mathbf{J}_v + \unitvec{z} \times \nabla
  \left( \frac{\lambda}{2} E^2 + \eta \right).
\end{equation}
The correlations of the noise $\eta$ are the same as in the original KPZ
equation. Indeed, it is straightforward to check that in the absence of
vortices, i.e., for $n_v = \mathbf{J}_v = 0$, the set of Maxwell
equations~\eqref{eq:divB},~\eqref{eq:Faraday},~\eqref{eq:Gauss},
and~\eqref{eq:Ampere} reduce to the non-compact KPZ equation if the electric
field is expressed in terms of the current as in Eq.~\eqref{eq:Villain-30} and
the latter is identified with $\mathbf{j} = \nabla \theta$.

The last $\delta$-functional in Eq.~\eqref{eq:7} encodes the equation of motion
of the vortices,
\begin{equation}
  \label{eq:vortex-Langevin}
  \frac{d \mathbf{x}_{\alpha}}{d t} = \mu n_{\alpha} \mathbf{E}(\mathbf{x}_{\alpha}) +
  \boldsymbol{\xi}_{\alpha}.
\end{equation}
Here, $\boldsymbol{\xi}_{\alpha}$ is a Markovian noise source, which is
introduced in the course of the HS decoupling of the noise vertex in
Eq.~\eqref{eq:11}, with correlations
\begin{equation}
  \label{eq:temp}
  \langle \xi_{\alpha i}(t) \xi_{\beta j}(t') \rangle = 2 \mu T
  \delta_{\alpha\beta} \delta_{ij} \delta(t - t').
\end{equation}
This completes the derivation of the dual electrodynamic theory.

\section{Outlook}
\label{sec:outlook}

Apart from the specific application to the compact KPZ equation in two spatial
dimensions, a promising future direction is to generalize the duality
transformation developed in this paper to treat other models of stochastic in-
or out-of-equilibrium dynamics of compact fields. Even the most straightforward
generalization to the one-dimensional compact KPZ equation should make it
possible to study the influence of phase slips on the scaling properties of
driven-dissipative condensates~\cite{He2015,Gladilin2014,Ji2015}. Moreover, it
will be interesting to see whether the same methods can be extended to quantum
systems which are described by Keldysh functional
integrals~\cite{Kamenev2011,Altland2010a}, and thus allow us to study real-time
dynamics of compact fields also in this case.

Finally, we note that the Coulomb gas picture of the static $XY$-model is the
starting point for deriving another representation in terms of a sine-Gordon
field theory~\cite{Minnhagen1987}. The advantage of this form is that it is
amenable to standard field theoretic tools and renormalization procedures. An
interesting question is whether a similar mapping exists in the context of the
compact KPZ equation.

\section*{Acknowledgments}
\label{sec:acknowledgments}

We are grateful to U. C. T\"auber and L. He for helpful
discussions. L. M. S. acknowledges support from the Koshland fellowship at the
Weizmann Institute of Sciences. L. M. S. and E. A. acknowledge further support
from the European Research Council through Synergy Grant UQUAM. G. W. and
E. A. acknowledge support from ISF under grant number 1594-11. G. W. was
additionally supported by the NSERC of Canada, the Canadian Institute for
Advanced Research, and the Center for Quantum Materials at the University of
Toronto. S. D. acknowledges funding by the German Research Foundation (DFG)
through the Institutional Strategy of the University of Cologne within the
German Excellence Initiative (ZUK 81), and by the European Research Council via
ERC Grant Agreement n. 647434 (DOQS).

\appendix

\section{Villain approximation}
\label{sec:vill-appr}

Here we compare the Villain form Eq.~\eqref{eq:Villain-18} to the standard
Villain approximation~\cite{Villain1975,Chaikin1995} for the static
$XY$-model. Hence, we begin by briefly reviewing the Villain approximation
for the latter case.

In view of performing a Villain-type approximation, the main difference between
the partition function for the classical $XY$-model and the MSR partition
function for the compact KPZ equation is that in the former case the weight of a
specific configuration is the \emph{real} Boltzmann factor, whereas in the
latter case we have to deal with a \emph{complex} weight (the last factor on the
RHS in Eq.~\eqref{eq:Villain-15}). For the classical $XY$-model, the Boltzmann factor
can be expanded in a Fourier series as
\begin{equation}
  \label{eq:Villain-12}
  e^{K \left( \cos(\theta) - 1 \right)} = \sum_{n = - \infty}^{\infty} e^{i n
    \theta + V(n)},
\end{equation}
where the potential $V(n)$ can be expressed in terms of the modified Bessel
function of the first kind $I_n(K)$,
\begin{equation}
  \label{eq:Villain-14}
  e^{V(n)} = \frac{1}{2 \pi} \int_0^{2 \pi} d \theta e^{-i n \theta + K
    \left( \cos(\theta) - 1 \right)} = e^{-K} I_n(K).
\end{equation}
This relation holds for any $K \in \C$. In the $XY$-model, the prefactor $K$ is
proportional to the inverse temperature, $K \propto 1/T$.
Usually~\cite{Chaikin1995} it is argued that the Villain approximation replaces
the exact potential $V(n)$ in Eq.~\eqref{eq:Villain-14} by an expression that
(i) is asymptotically equivalent in the low-temperature limit $K \to \infty$
(but see also~\cite{Janke1986}); moreover, obviously it should be possible to
(ii) interpret the approximate potential as a valid free energy functional for
$n$ which (iii) is computationally simpler to handle than the exact
expression. The crucial point is that by expanding the Boltzmann factor in a
Fourier series in Eq.~\eqref{eq:Villain-12}, it is guaranteed that for
\emph{any} approximation to $V(n)$ we still get a periodic function in $\theta$,
i.e., the compactness of $\theta$ is properly accounted for. The usual choice is
\begin{equation}
  \label{eq:Villain-72}
  e^{V(n)} \approx \frac{1}{\sqrt{2 \pi K}} e^{-n^2/(2 K)}.
\end{equation}
Obviously, this satisfies the criteria (ii) and (iii) formulated above. To see
in which sense the criterion (i) is satisfied, let us compare the asymptotic
expansions of Eqs.~\eqref{eq:Villain-14} and~\eqref{eq:Villain-72} for
$K \to \infty$~\cite{Abramowitz1964}: for the modified Bessel function we have
\begin{multline}
  \label{eq:Villain-71}
  e^{-K} I_n(K) \sim \frac{1}{\sqrt{2 \pi K}} \left[ 1 - \frac{4 n^2 - 1}{8 K}
  \right. \\ \left. + \frac{\left( 4 n^2 - 1 \right) \left( 4 n^2 - 9
      \right)}{2!  \left( 8 K \right)^2} + \dotsb \right],
\end{multline}
and the expansion of the Villain potential reads
\begin{equation}
  \label{eq:Villain-74}
  \frac{1}{\sqrt{2 \pi K}} e^{-n^2/(2 K)} \sim \frac{1}{\sqrt{2 \pi K}} \left[
    1 - \frac{4 n^2}{8 K} + \frac{\left( 4 n^2 \right)^2}{2! \left( 8 K
      \right)^2} + \dotsb \right].
\end{equation}
The pattern (which persists to higher orders) is, that of the polynomials
appearing in the numerators in Eq.~\eqref{eq:Villain-71}, in the Villain approximation
only the highest order terms in $n$ are kept.

When going from the static case of Eq.~\eqref{eq:Villain-12} to the dynamical
one in Eq.~\eqref{eq:Villain-17}, $K$ is replaced by $i K_{\sigma X}$, where in
the latter case $K_{\sigma X}$, which is defined in Eq.~\eqref{eq:Villain-16},
is real but can be both positive and negative --- depending in the equilibrium
case only on $\sigma$ and for $\lambda \neq 0$ also on the values of
$\epsilon, \lambda,$ and $\tilde{n}_{v X}$ --- and there is no reason to assume that its
absolute value is large. Therefore, the requirement (i) formulated above does
not apply. As with regard to the condition (ii), we should now request that the
Villain form gives a sensible contribution to the MSR action, i.e., it should
obey causality and lead to excitations which are stable (not growing in time),
which is not guaranteed \textit{a priori}. This requirement and also the
requirement of simplicity (iii) are met for the choice in
Eq.~\eqref{eq:Villain-18}, that results from replacing $K \to i K_{\sigma X}$ in
Eq.~\eqref{eq:Villain-12}.

\begin{widetext}
For completeness, let us briefly point out what happens to condition (i) in the
dynamical case. Upon replacing $K$ by $i K$, the exact Fourier
coefficient~\eqref{eq:Villain-14}, becomes
\begin{equation}
  \label{eq:Villain-75}  
  e^{i V(n)} = \frac{1}{2 \pi} \int_0^{2 \pi} d \theta e^{-i n \theta + i K
    \left( \cos(\theta) - 1 \right)} = e^{-i K} I_n(i K) = e^{-i \left(
      K - n \pi/2 \right)} J_n(K),
\end{equation}
where $J_n(K)$ is the Bessel function of the first kind. Then, the asymptotic
expansions analogous to Eqs.~\eqref{eq:Villain-71} and~\eqref{eq:Villain-74}
read
\begin{multline}
  \label{eq:Villain-73}
  e^{-i \left( K - n \pi/2 \right)} J_n(K) \sim \frac{1}{\sqrt{2 \pi K}} \left[
    1 + \frac{i \left( 4 n^2 - 1 \right)}{8 K} - \frac{\left( 4 n^2 - 1 \right)
      \left( 4 n^2 - 9 \right)}{2!
      \left( 8 K \right)^2} + \dotsb \right] \\
  + \frac{i e^{-i 2 \left( K - n \pi/2 \right)}}{\sqrt{2 \pi K}} \left[ 1 -
    \frac{i \left( 4 n^2 - 1 \right)}{8 K} - \frac{\left( 4 n^2 - 1 \right)
      \left( 4 n^2 - 9 \right)}{2!  \left( 8 K \right)^2} + \dotsb\right],
\end{multline}
and
\begin{equation}
\label{eq:Villain-77}
\frac{1}{\sqrt{i 2 \pi K}} e^{i n^2/(2 K)} \sim \frac{1}{\sqrt{2 \pi K}}
\left[ 1 + \frac{i 4 n^2}{8 K} - \frac{\left( 4 n^2 \right)^2}{2! \left( 8 K
    \right)^2} + \dotsb \right].
\end{equation}
\end{widetext}
The crucial difference to the static case is that the prefactor of the second
term in Eq.~\eqref{eq:Villain-73} is oscillating and not exponentially decaying,
and therefore it gives a contribution in the large-$K$ limit. Hence, asymptotic
equivalence cannot be used as an argument to motivate replacing the exact
Fourier coefficient~\eqref{eq:Villain-73} by the Villain
form~\eqref{eq:Villain-77}.

\bibliography{KPZ_Villain}

\end{document}